\documentclass[10pt, a4paper, twocolumn, twoside]{article} 
\usepackage{graphicx}
\usepackage{amsfonts, amsmath, amsthm, amssymb}
\usepackage{wrapfig}
\usepackage{hyperref}
\usepackage[english]{babel} 

\usepackage{microtype} 

\usepackage{amsmath,amsfonts,amsthm} 

\usepackage[svgnames]{xcolor} 

\usepackage[small, labelfont=bf, up]{caption} 

\usepackage{booktabs} 

\usepackage{lastpage} 

\usepackage{graphicx} 

\usepackage{enumitem} 
\setlist{noitemsep} 

\usepackage{sectsty} 
\allsectionsfont{\usefont{OT1}{phv}{b}{n}} 

\usepackage{geometry} 

\usepackage[T1]{fontenc} 
\usepackage[utf8]{inputenc} 

\usepackage{XCharter} 

\usepackage{hyperref}

\usepackage{journals}

\usepackage{fancyhdr} 
\newcommand{\shorttitle}[1]{\fancyhead[CE]{\textsl{#1}}}
\newcommand{\shortauthors}[1]{\fancyhead[CO]{\textsl{#1}}}

\fancypagestyle{firstpage}{ 
	\fancyhf{}
        \lhead{\vspace*{0.0em} \textit{\footnotesize 21$^{\rm st}$ European Workshop on White
            Dwarfs \\ Castanheira, Campos, Montgomery, eds. \\[-0.2em]
          July 23--27, 2018, Austin, Texas, USA}}
}

\date{}

\newcommand{\authorstyle}[1]{{\large\usefont{OT1}{phv}{b}{n}\color{DarkRed}#1}} 

\newcommand{\institution}[1]{{\footnotesize\usefont{OT1}{phv}{m}{sl}\color{Black}#1}} 

\usepackage{titling} 

\newcommand{\HorRule}{\color{DarkGoldenrod}\rule{\linewidth}{1pt}} 

\pretitle{
	\vspace{-30pt} 
	\HorRule\vspace{10pt} 
	\fontsize{16}{12}\usefont{OT1}{phv}{b}{n}\selectfont 
	\color{DarkRed} 
}

\posttitle{\par\vskip 10pt} 

\preauthor{} 

\postauthor{ 
	\vspace{0pt} 
	{\par\HorRule} 
	\vspace{-10pt} 
}

\newcommand{\newabstract}[1]{
    {\section*{Abstract}
    \bfseries #1}
  }

\usepackage[authoryear]{natbib}
\title{Evolutionary and pulsational properties of 
ultra-massive white dwarfs.
 The role of oxygen-neon phase separation.} 

\shorttitle{Evolutionary and pulsational properties of 
ultra-massive white dwarfs.} 

\shortauthors{De Gerónimo et al.} 

\author{
        \authorstyle{F.~C.~De Ger\'onimo,$^{1,2}$, M.~E.~Camisassa$^{1,2}$, A.~H.~C\'orsico$^{1,2}$, L.~G.~Althaus,$^{1,2}$}
	\newline\newline 
        $^1$\institution{Facultad de Ciencias Astron\'omicas y Geof\'isicas, Universidad
          Nacional de La Plata, Paseo del Bosque s/n, (1900) La Plata, Argentina; fdegeronimo@fcaglp.unlp.edu.ar}\\ 
	$^2$\institution{Instituto de Astrof\'isica  La Plata, CONICET-UNLP}\\ 
      }

\begin{document}

\maketitle 

\thispagestyle{firstpage} 


\newabstract{Ultra-massive hydrogen-rich white dwarfs (WDs) stars are
  expected to harbor oxygen/neon cores resulting from semi-degenerate
  carbon burning when the progenitor star evolves through the super
  asymptotic giant branch (SAGB) phase.  These stars are expected to
  be crystallized by the time they reach the ZZ Ceti domain. We show
  that crystallization leads to a phase separation of oxygen and neon
  in the core of ultra-massive WDs, which impacts markedly
  the pulsational properties, thus offering a unique opportunity to
  study the processes of crystallization and to infer the core
  chemical composition in WD stars.}


\section{Methodology \& input physics}
We computed the evolution \citep{2018arXiv180703894C} and pulsation
properties \citep{2018arXiv180703810D} of ultra-massive DA
(hydrogen-rich) WD sequences with stellar masses
$M_{\star}= 1.10, 1.16, 1.22$, and $1.29 M_{\odot}$ (Fig. \ref{fig:g-teff})
resulting from the complete evolution of the progenitor stars through
the SAGB phase \citep{2010A&A...512A..10S}.  Prior evolution provides
us with realistic core chemical profiles, envelope stratification and helium
mass. In table \ref{tabla1} we show the He mass of our models, together
with the effective temperature and surface gravity at the onset of
crystallization, and the fraction of crystallized mass at the blue and
red edges of the ZZ Ceti instability strip.  The cores are composed
mostly of $^{16}$O and $^{20}$Ne, and smaller amounts of $^{12}$C,
$^{23}$Na and $^{24}$Mg.  The H content is set to
$M_{\rm H} \sim 10^{-6} M_\star$.

Nonradial $g$(gravity)-mode pulsations of our complete set of
ultra-massive ONe-core DA WD models were computed using the adiabatic
version of the {\tt LP-PUL} pulsation code \citep{2005A&A...429..277C}.
The pulsation code is
based on the general Newton-Raphson technique that solves the full
fourth order set of equations and boundary conditions governing
linear, spheroidal, adiabatic, non-radial stellar pulsations following
the dimensionless formulation of \citet{1971AcA....21..289D}. To account for
the effects of crystallization on the pulsation spectrum of $g$ modes,
we adopted the “hard sphere” boundary conditions, which assume that
the amplitude of the eigenfunctions of $g$ modes is drastically reduced
below the solid/liquid interface due to the non-zero shear modulus of the
solid, as compared with the amplitude in the fluid region
\citep[see][]{1999ApJ...526..976M}.

The DA WD evolutionary models developed in this work were computed
with the amply used {\tt LPCODE} evolutionary code
\citep{2005A&A...435..631A}. This code considers all the physical
ingredients involved in WD, including element diffusion.
In this work we included, \emph{for the first time}, both
energy release and chemical abundance changes caused by the
process of phase separation during
crystallization.  We considered the phase diagram of
\citet{2010PhRvE..81c6107M}, suitable for the dense plasma of
oxygen/neon mixtures appropriate for ultra-massive WDs.
We computed an additional sequence of CO-core WDs with
$M_{\star}= 1.10 M_{\odot}$, for which crystallization was considered
following the phase diagram of \citet{2010PhRvL.104w1101H}.
Phase diagrams provide us with the temperature at which crystallization
occurs and the resulting abundances of the solid phase.  To our knowledge,
this is the \emph{first} evolutionary and pulsational analysis of
ultra-massive DA WD models that includes phase separation processes in
ONe cores.


\section{Results}

\begin{figure}
\includegraphics[width=1\linewidth]{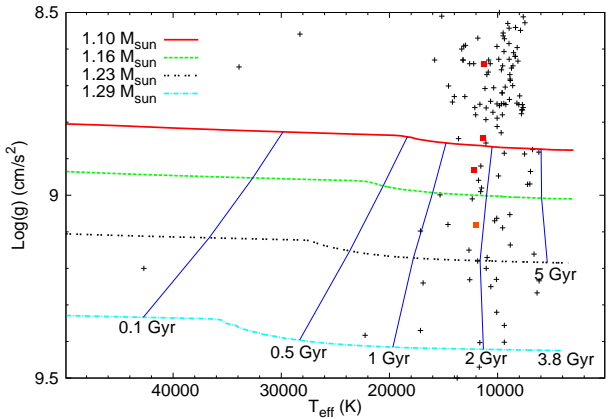}
\caption{Hydrogen-rich WD model sequences in the $\log g -T_{\rm eff}$
  plane.  Blue solid lines display isochrones of 0.1, 0.5, 1, 2, and 5 Gyr. 
Plus symbols indicate the location of a sample of ultra-massive DA WDs.
Squares display ultra-massive ZZ Ceti stars.}
\label{fig:g-teff}
\end{figure}

\begin{figure}
\includegraphics[width=1\linewidth]{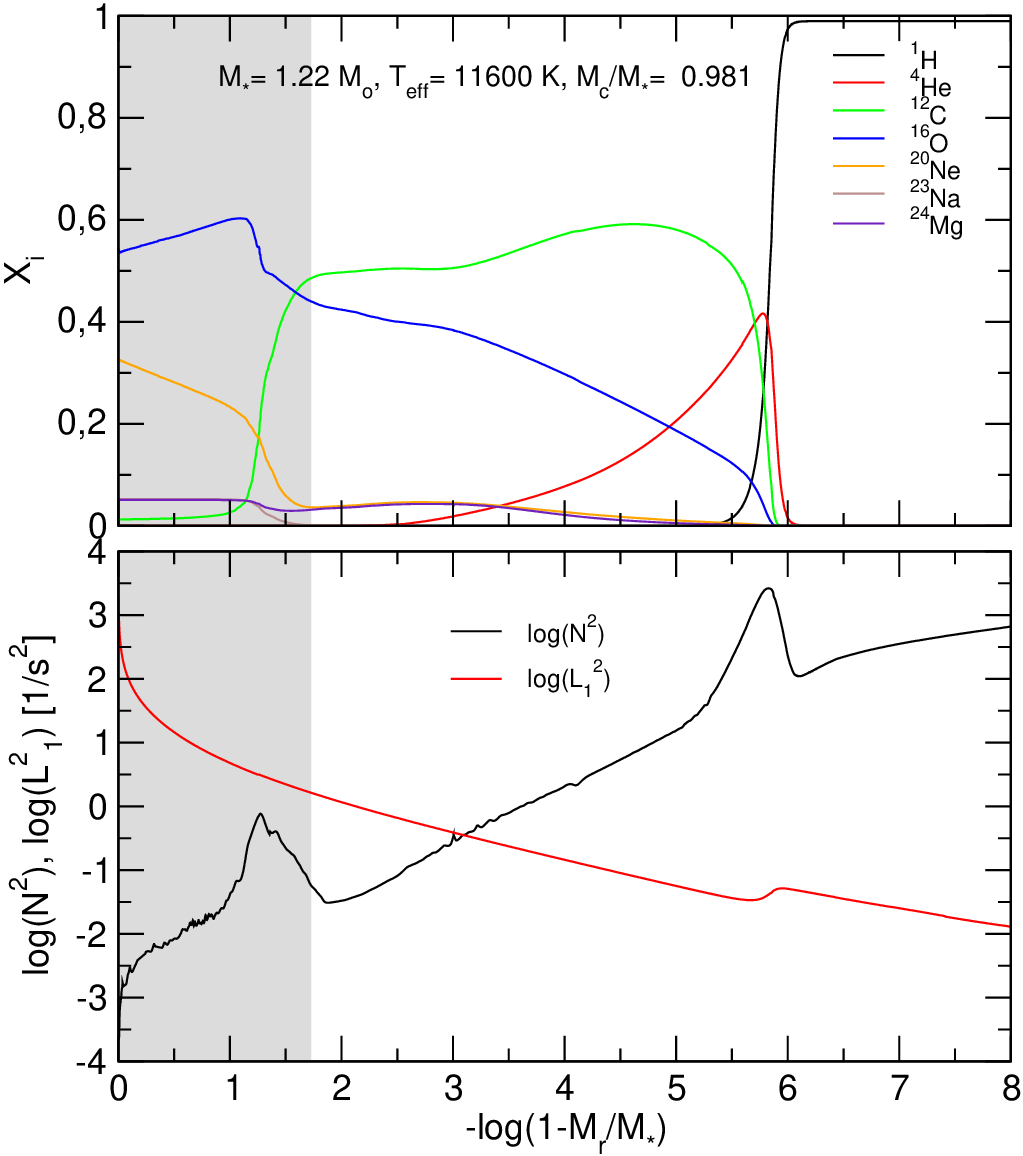}
\captionof{figure}{Chemical profiles (upper panel), and the logarithm
  of the squared Brunt-V\"ais\"al\"a and Lamb ($\ell= 1$)
  frequencies (lower panel), corresponding to a ONe-core WD model
  with $M_{\star}= 1.22 M_{\odot}$ and $T_{\rm eff} \sim 11\,600$ K.
  The gray area marks the domain of crystallization.
  $M_c/M_{\star}$ is the crystallized mass fraction of the model.}
\label{fig:X-B-BVF-122-11500}
\end{figure}

\begin{table*}[]
\centering

\begin{tabular}{lccccc}
\hline
  $M_{\star}/M_{\odot}$   &$M_{\rm He}/M_{\star}$    & $T_{\rm eff}^{\rm c}$&$\log g^{\rm c}$& $M_{\rm c}/M_{\star}$ & $M_{\rm c}/M_{\star}$ \\
 & ($\times 10^{-5}$) & (K) & (cgs) & ($T_{\rm eff}= 12\,500$ K) & ($T_{\rm eff}= 10\,500$ K) \\
\hline
  1.098          & 29.6 & 19881  & 8.83     & 0.81                            &0.92                             \\
  1.159           & 15.7 & 23291  & 8.95     & 0.90                            &0.96                                \\
  1.226          & 6.38 & 28425  & 9.12     & 0.96                            &0.98                                 \\
  1.292          & 1.66 & 37309  & 9.33     & 0.994                           &0.998                                \\
\hline
\end{tabular}
\caption{He mass content of our ONe-core ultra-massive DA WD models,
  together with the effective temperature and surface gravity at the
  onset of crystallization, and the fraction of crystallized mass at
  the blue and red edges of the ZZ Ceti instability strip.}
\label{tabla1}
\end{table*}

In Fig. \ref{fig:g-teff} we show the $\log g - T_{\rm eff}$ plane of
our WD sequences, computed for this work. In addition, we plot the
isochrones of 0.1, 0.5, 1, 2 and 5 Gyr (blue lines) together with the
sample of ultra-massive DA WDs known at date (black plus symbols) and
ultra-massive ZZ Ceti stars (red squares). It can be seen a change of
slope of the isochrones, reflecting the dependence of cooling times on
the mass of the WD: at early stages, evolution proceeds slower in more
massive WDs, while the opposite trend is found at advanced stages.

We found that phase separation during crystallization strongly modifies
the core chemical profiles of our models, and this has a no-negligible
impact on the period spectrum. Fig.  \ref{fig:X-B-BVF-122-11500} displays
the chemical profiles and the
logarithm of the squared
Brunt-V\"ais\"al\"a and Lamb ($\ell= 1$) frequencies in terms of the
outer mass fraction corresponding to a ONe-core WD model with
$M_{\star}= 1.16M_{\odot}$ and  $T_{\rm eff} \sim 11\,600$ K and a
percentage of crystallized mass of 98 \%. Note that for this model,
phase separation of $^{16}$O and $^{20}$Ne has already finished, and
the triple chemical transition region C/O/Ne is located at 
$-\log(1-M_r/M_{\star}) \sim 1.3$, well within the crystallized region.
By virtue of this, the
bump in the Brunt-V\"ais\"al\"a frequency associated to this chemical
interface (see lower panel) is completely irrelevant for the pulsation
properties of the model, since the $g$ modes cannot propagate in the
crystallized core. This implies that the only relevant chemical
interface of the model is that of the $^{1}$H and $^{4}$He, located at
$-\log(1-M_r/M_{\star}) \sim 5.8$. 

\begin{figure}
\includegraphics[width=1\linewidth]{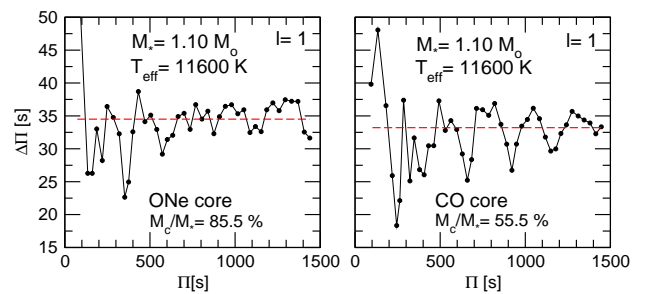}
\captionof{figure}{Forward period spacing ($\Delta \Pi$) in terms of
  the periods of $\ell= 1$ pulsation $g$ modes for $1.10 M_{\odot}$
  WD models at $T_{\rm eff} \sim 11600$ K with an  ONe core (left
  panel) and a CO core (right panel). The
  percentages of the crystallized mass are indicated. The horizontal
  red-dashed line is the asymptotic period spacing.}
\label{fig:delp-110-ONE-CO}
\end{figure}

To explore the possibility of find out the core composition of ultra-massive
ZZ Ceti stars through their pulsations, we have compared the period-spacing
diagrams of CO-core and ONe-core WD
models with  $1.10 M_{\odot}$ and the same effective temperature.
In Fig. \ref{fig:delp-110-ONE-CO} we depict the period spacing ($\Delta \Pi$) in
terms of the periods of $\ell= 1$
pulsation $g$ modes for two WD models at $T_{\rm eff} \sim 11600$ K with an
ONe core (left) and a CO core (right). In both models, latent-heat
release and chemical redistribution caused by phase separation have
been taken into account during crystallization. We found that period-spacing
departures from the mean period separation
due to mode-trapping effects are smaller for the ONe-core  model than
for the CO-core model.  We conclude that the features found in the
period-spacing diagrams could be employed to differentiate the chemical
composition of the cores of ultra-massive ZZ Ceti stars. 

\section{Conclusions}

We found that phase separation during crystallization in the core of
ultra-massive DA WDs have a marked impact on the pulsation properties
of ultra-massive ZZ Ceti stars. Interestingly enough, we found that
the differences in the period-spacing diagrams could be employed 
to infer the core composition of ultra-massive ZZ Ceti stars, something
that should be complemented with detailed asteroseismic analyses using the
individual observed periods. Promising targets to to achieve this goal
are the  ZZ Ceti stars BPM 37093
\citep{1992ApJ...390L..89K,2005A&A...432..219K}, GD 518
\citep{2013ApJ...771L...2H}, and SDSS J0840+5222 \citep{2017MNRAS.468..239C}.

\section*{Acknowledgments} F.C.D.G and A.H.C. warmly thank
the Local Organizing Committee of the  21th European White Dwarf
Workshop for support that allowed him to attend this conference.

\vspace*{0.5em}

The evolutionary sequences presented in this work can be found at:
\url{http://evolgroup.fcaglp.unlp.edu.ar/TRACKS/DA.html}


\bibliography{papers}

\end{document}